\documentstyle[preprint,tighten,prl,aps]{revtex}
\begin{document}

\newcommand{\be}{\begin{equation}}
\newcommand{\bestar}{\[\extraspace}
\newcommand{\ee}{\end{equation}}
\newcommand{\eestar}{\]}
\newcommand{\bea}{\begin{eqnarray}}
\newcommand{\eea}{\end{eqnarray}}

\newcommand{\nonu}{\nonumber \\[2mm]}
\newcommand{\ie}{{\it i.e.}\ }

\newcommand{\half}{\frac{1}{2}}
\newcommand{\si}{\sigma}

\newcommand{\cd}{c^{\dagger}}
\newcommand{\Sd}{S^{\dagger}}
\newcommand{\Sz}{S^z}
\newcommand{\etad}{\eta^{\dagger}}
\newcommand{\etaz}{\eta^z}

\newcommand{\up}{\uparrow}
\newcommand{\down}{\downarrow}

\newcommand{\vac}{| \, 0 \, \rangle}
\newcommand{\avac}{\langle \, 0 \, |}
\newcommand{\sumnn}{\sum_{\langle jl \rangle}}
\newcommand{\grst}{|\,\psi_N\,\rangle}

\newcommand{\np}{Nucl.\ Phys.\ }
\newcommand{\phl}{Phys.\ Lett.\ }
\newcommand{\cmp}{Comm.\ Math.\ Phys.\ }
\newcommand{\pr}{Phys.\ Rev.\ }
\newcommand{\phrl}{Phys.\ Rev.\ Lett.\ }
\newcommand{\jop}{J.\ Phys.\ }

\draft

\preprint{ITP-SB-94-18}
\title{$\eta$-pairing as a mechanism of superconductivity in models of
strongly correlated electrons}

\author{Jan de Boer\footnote{email: {\sc deboer@max.physics.sunysb.edu}},
Vladimir E.~Korepin\footnote{email: {\sc korepin@max.physics.sunysb.edu}}
and
Andreas Schadschneider\footnote{email: {\sc aceman@max.physics.sunysb.edu}}
}
\address{Institute for Theoretical Physics\\
State University of New York at Stony Brook\\
Stony Brook, NY 11794-3840, U.S.A.}

\date{\today}
\maketitle
\begin{abstract}
We consider extended versions of the Hubbard model which contain additional
interactions between nearest neighbours. In this letter we
show that a large class of these models has a superconducting
ground state in arbitrary dimensions. In some special cases we are able to
find the complete phase diagram. The superconducting phase
exist even for moderate repulsive values of the Hubbard interaction $U$.
\end{abstract}

\narrowtext
In this letter we consider extensions of the Hubbard model. We shall show
that they are superconducting. The idea to explain superconductivity in the
framework of strongly correlated electrons was proposed in \cite{ARZ} and
has subsequently been studied in numerous publications. Most of these
investigations used approximate or numerical methods and so only a few
exact results are known. One of these exact results is an algebraic
approach based on so-called $\eta$-pairs \cite{pairing}. It allows for the
construction of states exhibiting off-diagonal-long-range-order
(ODLRO). This important concept has been developed in \cite{odlro}.
ODLRO is an adequate definition of superconductivity in arbitrary dimensions
since it implies also the Meissner effect and flux quantisation
\cite{odlro,sewell,nieh}.

Let us first consider the Hamiltonian of the Hubbard model
\be
{\cal H}_0(U) = -t \sumnn \sum_{\sigma =\up\down}\left( \cd_{j\sigma}
c_{l\sigma} + \cd_{l\sigma}c_{j\sigma} \right)
+ U \, \sum_{j=1}^L (n_{j\up}-\half)(n_{j\down}-\half)\ .
\label{hubb}
\ee
Here $c_{j\sigma}$ are canonical Fermi operators which describe electrons
on a $d$-dimensional lattice, i.e.
$\{ c^\dagger_{j\sigma} , c_{l\tau} \} = \delta_{jl}\delta_{\sigma\tau}$ and
$c_{j\sigma} \vac = 0$
where $\vac $ denotes the Fock vacuum.

In (\ref{hubb}) $j$ runs through all $L$ sites of the $d$-dimensional lattice,
$\langle jl\rangle$ denotes nearest-neighbour sites and $U$ is the
Hubbard coupling. $n_{j\si}= c^\dagger_{j\si} c_{j\si}$ denotes
the number operator for electrons with spin $\si$ on site
$j$ and we write $n_j=n_{j\up} + n_{j\down}$.

\vskip 3mm

This letter consists of two parts. In the first part we shall add
general nearest-neighbour interactions to the Hubbard Hamiltonian
(\ref{hubb}) and analyze the three following questions:
\begin{itemize}
\item When does this new Hamiltonian commute with the $\eta$-operators
\be
\eta = \sum_{j} c_{j\up}c_{j\down}\ , \ \ \
\eta^\dagger = \sum_{j} \cd_{j\down}\cd_{j\up} \ .
\label{eta}
\ee
\item When will $\left(\eta^\dagger\right)^N\vac $ be an exact
eigenstate of this Hamiltonian?
\item When will $\left(\eta^\dagger\right)^N\vac $ be a ground state
of this Hamiltonian?
\end{itemize}

We shall answer all these questions for the general multiparametric
Hamiltonian. For example, $\grst=\left(\eta^\dagger\right)^N\vac $ will be
a ground state in a 9-parametric region (subject to some inequality).

What makes the state $\grst$ special is the fact that it has been shown
to have ODLRO (and thus is superconducting),
\be
\langle \psi_N | \cd_{j\down} \cd_{j\up}
c_{l\up} c_{l\down} | \psi_N \rangle \quad \stackrel{|l-j| \to \infty
}{\longrightarrow} \quad \frac{N}{L} (1-\frac{N}{L}) \ ,
\label{odlro0}
\ee
where we have also taken the thermodynamic limit ($N,L \to \infty$ with
$N/L$ fixed).

\vskip 3mm

The results of the first part show that superconductivity based on
$\eta$-pairs is a rather typical phenomenon. To see the relation
between $\eta$-pairs and the usual order parameter of BCS theory
we express the $\etad$-operator in terms of electronic
operators $\cd_{k\si}$ in momentum space ,
$\cd_{j\si} ={1 \over \sqrt{L}}\sum_k e^{ijk}\cd_{k\si}$,
which yields $\etad = \sum_k \cd_{k\down}\cd_{-k\up}$. This is just
the BCS order parameter.

\vskip 3mm

In the second part of this letter we will consider two special
cases of the general Hamiltonian discussed in the first part. Both of these
models are exactly solvable in one dimension and have $\eta$-pairs in
the ground state (and thus are superconducting) even for moderate repulsive
values of $U$. These results are valid in any dimension.

\vskip 3mm

In order to argue that superconductivity based on $\eta$-pairing is a
generic rather than an exotic phenomenon, we consider the Hamiltonian
${\cal H}(U)={\cal H}_0 (U) + {\cal H}_1$, where ${\cal H}_0 (U)$
is the Hubbard Hamiltonian (\ref{hubb}), and
\bea\label{gham}
{\cal H}_1 & = &
X\sum_{<jl>,\si} ( \cd_{j\si} c_{l\si} + \cd_{l\si} c_{j\si} )
(n_{j,-\si}+n_{l,-\si}) \nonu
& & +V\sum_{<jl>} (n_j-1)(n_l-1) + J_z \sum_{<jl>} S^z_j S^z_l \nonu
& & +\frac{J_{xy}}{2} \sum_{<jl>} (\Sd_j S_l + \Sd_l S_j) +
Y\sum_{<jl>} (\cd_{j\up} \cd_{j\down} c_{l\down} c_{l\up} +
 \cd_{l\up} \cd_{l\down} c_{j\down} c_{j\up}) \nonu
& & +P\sum_{<jl>} \left(
(n_{j\up}-\frac{1}{2}) (n_{j\down}-\frac{1}{2}) (n_l-1) +
(n_{l\up}-\frac{1}{2}) (n_{l\down}-\frac{1}{2}) (n_j-1) \right) \nonu
& & +Q \sum_{<jl>}
(n_{j\up}-\frac{1}{2}) (n_{j\down}-\frac{1}{2})
(n_{l\up}-\frac{1}{2}) (n_{l\down}-\frac{1}{2}), \nonu
& &
+\mu \sum_j n_j + h \sum_{j}(n_{j\up}-n_{j\down})
\eea
where the $SU(2)$ spin operators $S_j^a$ are given by $S_j^{z}=
\frac{1}{2}(n_{j\up}-n_{j\down})$, $S_j=c^{\dagger}_{j\down} c_{j\up}$
and $S_j^{\dagger}=c^{\dagger}_{j\up} c_{j\down}$.
The first term in (\ref{gham}) is known as the bond-charge interaction,
the second one is the nearest-neighbour Coulomb interaction. In addition
we included a $XXZ$-type spin interaction with exchange constants $J_{xy}$
and $J_z$ between nearest neighbour sites.
The relevance of the pair-hopping term $Y$ for high-temperature
superconductivity has recently been discussed in \cite{CSAS}.
Apart from a chemical potential $\mu$ and a magnetic field $h$ we also
added a three- and four-particle density interaction $P$ and $Q$, respectively.

The Hamiltonian (\ref{gham}) with $t=X$ and zero magnetic field
$h=0$ is the most general Hamiltonian that one can write down that
is hermitian, symmetric under spinflip ($\sigma \to -\sigma$) and conserves
the total number $N_\sigma =\sum_{j=1}^L n_{j\sigma}$ of electrons with spin
$\sigma$ and the number of doubly occupied sites. The last
requirement is quite natural if we consider exact $\eta$-pairing
ground states, since these are eigenstates of the number operator for
doubly occupied sites $N_2=\sum_j n_{j\up} n_{j\down} $. To include
the Hubbard model itself we allow $t \neq X$ and have for convenience
also included a chemical potential $\mu$ and allowed for nonzero magnetic
field $h$.

The conditions under which ${\cal H}(U)$ commutes with $\eta^{\dagger}$ can be
read of from the identity
\bea    \label{heta}
[{\cal H}(U),\eta^{\dagger}] & = & 2(t-X)\sum_{<jl>} (\cd_{l\up} \cd_{j\down}+
\cd_{j\up} \cd_{l\down} ) \nonu
& & + (2V-Y)\sum_{<jl>} (\eta^{\dagger}_j (n_l-1) +
 \eta^{\dagger}_l (n_j-1)) \nonu
& & + 2P \sum_{<jl>} (\eta^{\dagger}_j
(n_{l\up}-\frac{1}{2}) (n_{l\down}-\frac{1}{2})
+\eta^{\dagger}_l
(n_{j\up}-\frac{1}{2}) (n_{j\down}-\frac{1}{2})) \nonu
& & + 2\mu\sum_j \eta^{\dagger}_j\ .
\eea
Therefore, the condition for $\eta$-symmetry is that $t=X$, $2V-Y=0$,
$P=0$ and $\mu=0$. Using (\ref{heta}) one finds that $(\etad)^N
\vac $ is an eigenstate of ${\cal H}(U)$ if $t=X$ and $2V=Y$. The
corresponding eigenvalue is
\be
E_{N}=\frac{UL}{4}+\frac{ZL}{2} \left( V+\frac{P}{2}+\frac{Q}{16}\right)+
N\left( 2\mu+\frac{PZ}{2}\right),
\label{energ}
\ee
where $Z$ is the number of nearest neighbours of a lattice site.

Next, we want to determine under which conditions the $\eta$-pairing state
is actually the ground state of the theory. For this, we use the following
lower bound for the ground state energy. Let $\{\psi_{\alpha} \}$ be an
orthonormal basis of the Hilbert space, then the ground state energy
$E_0$ satisfies
\be
E_0 \geq \min_{\alpha} \left(
\langle  \psi_{\alpha} | {\cal H}(U) | \psi_{\alpha} \rangle
- \sum_{\beta \neq \alpha} |\langle \psi_{\beta} |
{\cal H}(U) | \psi_{\alpha} \rangle | \right).
\ee
This lower bound has previously been used by Ovchinnikov \cite{ovchin}
to analyze ferromagnetic ground states continuing the work of
Strack and Vollhardt \cite{voll}. By requiring that this lower bound
equals $E_N$ in (\ref{energ}) one can prove \cite{deboer} that
$(\etad)^N\vac$ is a ground state of ${\cal H}(U)$ if $V\leq 0$ and
\be \label{bound}
\frac{-U}{Z} \geq \max
\left( |P|+2\frac{|h|}{Z}+\frac{Q}{4} + 2|t|+ 2V,\
V-\frac{J_z}{4}+\frac{2|h|}{Z},\
V+\frac{J_z}{4}+|\frac{J_{xy}}{2}| \right).
\ee

This shows that if $t=X$ and $2V=Y\leq 0$, one can for every value of the
parameters in ${\cal H}$ always make the ground state superconducting,
by making the Hubbard coupling sufficiently small.

The results of \cite{lee} suggest that even if we relax the conditions $t=X$
and $2V=Y$, we still will have a superconducting ground state for
sufficiently small $U$. This is much harder to prove, since the explicit
form of the ground state is not known, but one might get some
indications from doing perturbation theory in $t-X$ or $2V-Y$.

One might object that negative $U$ does not represent the real
physical situation, since $U$ represents the Coulomb repulsion
between electrons and that should be positive. However, we shall see
in a moment that in two special cases the following happens: as soon
as $U$ becomes larger than the bound (\ref{bound}), the ground state
becomes $(\etad)^{N-n}|\,\psi_n\,\rangle$, where $|\,\psi_n\,\rangle$ is a
$2n$-electron state without doubly occupied sites. As $U$ increases,
so does $n$ until $n=N$. At this value of $U$ the theory will no longer
be superconducting. It is therefore quite likely that superconductivity
will persist for a somewhat larger range of values of $U$ than
given by (\ref{bound}).

\vskip 3mm

After these general considerations let us now look at two models more
closely. For these models the full phase diagram in arbitrary dimensions
can be obtained. Furthermore, in one dimension we are able to calculate
the complete spectrum. From now on we will set $t=1$ for convenience.

The first model is obtained by setting all parameters in (\ref{gham}) equal to
zero except for the bond-charge interaction $X$,
\be
{\cal H}(X,U)= - \sumnn \sum_{\sigma =\up\down} \left(\cd_{j\sigma}
c_{l\sigma} + \cd_{l\sigma} c_{j\sigma} \right)
\left(1-X\left( n_{j,-\sigma}+n_{l,-\sigma}\right)\right)
+ U \, \sum_{j=1}^L (n_{j\up}-\half)(n_{j\down}-\half)\ .
\label{hhirsch}
\ee
For general values of
$X$ this model has been discussed extensively by Hirsch \cite{hirsch}.
He argued that for large densities of electrons (low doping) the
bond-charge interaction leads to an attractive effective interaction
between the holes which may even create Cooper-pairs of holes. Indeed,
using a BCS-type mean-field theory he found a superconducting phase for
small hole concentrations.

Unfortunately, up to now even in one dimension there are almost no exact
results available although it has been shown that a simplified
version of Hirsch's Hamiltonian in one dimension indeed has a strong
tendency towards superconductivity \cite{bksz}.

In the following we consider the Hamiltonian for the special case $X=1$.
As mentioned above, at this point the number of doubly occupied sites is
conserved. Therefore the Hubbard interaction acts as chemical potential
for the doubly occupied sites. The dynamics become quite simple since the
local Hamiltonian permutes bosons (i.e.~empty sites and
doubly occupied sites) with fermions (i.e.~singly occupied sites) on
neighbouring sites but not bosons with bosons or fermions
with fermions. In the sector with no double-occupations the Hamiltonian
reduces to the well-known $U=\infty$ Hubbard model.

In one dimension this Hamiltonian can be solved exactly \cite{as} by
generalising the method applied in \cite{ci,kotrla} to the $U=\infty$
Hubbard model. The Hilbert space is divided into certain subspaces in
which the Hamiltonian can be mapped onto spinless fermions with
twisted boundary conditions. The twisting angle depends on the subspace
considered. In this way the complete spectrum can be obtained. The
ground state energy for a system of $N_1$ (single) electrons and $N_2$
doubly occupied sites at $U=0$ is given by
\be
E_0(N_1,N_2)=-2\ {\sin(N_1\pi/L)\over \sin\left(\pi/L\right)} \ .
\label{gsenergy}
\ee
Note that this energy is independent of $N_2$. In general, the ground
state is highly degenerate as is known for the $U=\infty$ Hubbard model.
After including the Hubbard interaction we can minimize this energy
for a given total number $N=N_1+2N_2$ of electrons. We find the phase diagram
shown in fig.~\ref{Fig1}. For $U\leq -4$ (sector I) the $\eta$-pairing state
$\left(\etad \right)^{N/2}\vac$ is a ground state in agreement with the
inequality (\ref{bound}). In sector II where $-4\leq U \leq U_c(N)=
-4\cos(N\pi/L)$ the ground
state is of the form $\left(\etad\right)^{N/2} |U=\infty,N_1\rangle$
where $|U=\infty,N_1\rangle$ is the ground state of the $U=\infty$
Hubbard model for $N_1$ electrons. Both these states exhibit ODLRO
showing the existence of a superconducting ground state even for
moderately positive values of $U$. For $U\geq U_c(N)$ the ground state
is that of the $U=\infty$ Hubbard model (for densities $D=N/L>1$ it is the
state obtained by particle-hole symmetry).

The phase diagram in higher dimensions looks quite similar although
we can not give an explicit expression for $U_c(N)$. From (\ref{bound})
we know that $\left(\etad \right)^{N/2}\vac$ is a ground state for
$U\leq -2Z$. The rest of the phase diagram can be obtained by an
argumentation analogous to that in \cite{EKSb}. The properties
necessary for this argumentation to hold are
\begin{itemize}
\item $\eta$-symmetry,
\item conservation of the number of doubly occupied sites,
\item for $U=0$ the ground state energy is independent of $N_2$.
\end{itemize}

\vskip 3mm
Another interesting special case of the general Hamiltonian (\ref{gham})
is the supersymmetric Hubbard model \cite{EKSa,EKSb}. The Hamiltonian
of this model is given by
\be
{\cal H}_{sH} = {\cal H}^0
+ U \, \sum_{j=1}^L (n_{j\up}-\half)(n_{j\down}-\half)
\label{eksham}
\ee
where ${\cal H}^0=- \sumnn \, H^0_{j,l}$, and
\bea
H^0_{j,l} &=&
\cd_{l\up} c_{j\up}(1-n_{j\down}-n_{l\down})
+ \cd_{j\up} c_{l\up}(1-n_{j\down}-n_{l\down})
\nonu
&& + \cd_{l\down} c_{j\down}(1-n_{j\up}-n_{l\up})
+ \cd_{j\down} c_{l\down}(1-n_{j\up}-n_{l\up})
\nonu
&& + \half (n_j - 1)(n_l - 1)
+ \cd_{j\up} \cd_{j\down} c_{l\down} c_{l\up}
+ c_{j\down} c_{j\up} \cd_{l\up} \cd_{l\down}
\nonu
&& - \half (n_{j\up}-n_{j\down})(n_{l\up}-n_{l\down})
- \cd_{j\down} c_{j\up} \cd_{l\up} c_{l\down}
- \cd_{j\up} c_{j\down} \cd_{l\down} c_{l\up}
\nonu
&& + (n_{j\up}-\half)(n_{j\down}-\half)
+ (n_{l\up}-\half)(n_{l\down}-\half) \ .
\label{hamil0jl}
\eea
This Hamiltonian corresponds to the choice $X=t=1$, $V=-1/2$, $J_{xy} =
J_z = 2$, $Y=-1$ and $P=Q=\mu=h=0$ in (\ref{gham}). Note that $H^0_{j,l}$
also contains a Coulomb interaction term $-Z(n_{j\up}-\half)
(n_{j\down}-\half)$. ${\cal H}^0$ commutes with $\eta$ and $\etad$ as
defined in (\ref{eta}). Actually, the model has a larger set of  symmetries
that form the superalgebra  $U(2|2)$ \cite{EKSa}.

The ground state phase diagram has been obtained in \cite{EKSb}.
It looks very similar to fig.~\ref{Fig1}. The main difference is the
occurence of the ground state $|t-J\rangle$ of the supersymmetric
$t-J$ model \cite{tJ} that replaces $|U=\infty \rangle$ in fig.~\ref{Fig1}.
In addition, the sector I exists for all $U<0$ (in
every dimension) as can also be seen from the inequality (\ref{bound}).
A superconducting phase exist for moderate positive $U$ in all
dimensions. The additional interactions of (\ref{hamil0jl}) lift the
degeneracies encountered in the model (\ref{hhirsch}).
A more detailed discussion of the phase diagram of the supersymmetric
Hubbard model has been given in \cite{EKSb}.
\vskip 2mm

In conclusion, in this letter we have shown that $\eta$-pairs
provide a simple and clear mechanism of superconductivity.

\vskip 4mm
AS gratefully acknowledges financial support by the Deutsche
Forschungsgemeinschaft. JdB is sponsored in part by NSF grant
No.~PHY-9309888.

\begin{figure}
\caption{Phase diagram for the Hamiltonian (9) in one
dimension. $D=N/L$ is the particle density and $U$ the Hubbard interaction.
Qualitatively the phase diagram in higher dimensions has the same form. }
\label{Fig1}
\end{figure}
\vskip 17 true cm
\end{document}